\documentclass{ws-procs9x6}

\begin{document}

\title{
EQUATION OF STATE FOR DARK ENERGY IN MODIFIED GRAVITY THEORIES
}

\author{Kazuharu Bamba$^*$}

\address{Kobayashi-Maskawa Institute for the Origin of Particles and the
Universe, Nagoya University, \\
Nagoya 464-8602, Japan\\
$^*$E-mail: bamba@kmi.nagoya-u.ac.jp
}

\begin{abstract}
We explore the equation of state (EoS) for dark energy $w_{\mathrm{DE}}$ in 
modified gravitational theories to explain 
the current accelerated expansion of the universe. 
We explicitly demonstrate that the future crossings of the phantom 
divide line of $w_{\mathrm{DE}}=-1$ are the generic feature in the existing 
viable $f(R)$ gravity models. 
Furthermore, we show that the crossing of the phantom divide 
can be realized in the combined $f(T)$ theory constructed with 
the exponential and logarithmic terms. 
In addition, we investigate the effective EoS for the universe 
when the finite-time future singularities occur in non-local gravity. 
\end{abstract}


\bodymatter

\section{Introduction}\label{aba:sec1} 

It has been suggested that the current expansion of the universe 
is accelerating by recent cosmological observations such as 
Supernovae Ia (SNe Ia)\cite{SN1}, 
cosmic microwave background (CMB) radiation\cite{WMAP, Komatsu:2010fb}, 
large scale structure (LSS)\cite{LSS}, 
baryon acoustic oscillations (BAO)\cite{Eisenstein:2005su}, 
and weak lensing~\cite{Jain:2003tba}. 
There are two representative approaches to account for 
the late time cosmic acceleration. 
One is the introduction of cosmological constant dark energy 
in the framework of general relativity. 
The other is the modification of gravity, 
for example, $f(R)$ gravity, 
where $f(R)$ is an arbitrary function of the scalar curvature $R$ 
(for recent reviews on $f(R)$ gravity, 
see, e.g., Refs.~\refcite{Review-Nojiri-Odintsov, 
Book-Capozziello-Faraoni}). 
One of the most important parameter in this issue is 
the equation of state (EoS) for dark energy 
$w_{\mathrm{DE}} \equiv P_{\mathrm{DE}}/ \rho_{\mathrm{DE}}$, 
which is the ratio of the pressure $P_{\mathrm{DE}}$ of dark energy to 
the energy density $\rho_{\mathrm{DE}}$ of it. 
Recent cosmological observational data\cite{observational-status} also seems 
to indicate the crossing of the phantom divide line of 
$w_{\mathrm{DE}}=-1$ of the EoS for dark energy in the near ``past''. 
In this paper, we concentrate on the evolution of $w_{\mathrm{DE}}$. 
In particular, we review our main results on it in 
$f(R)$ gravity\cite{Bamba:2010iy}, $f(T)$ theory\cite{
BGLL-BGL
} and non-local gravity\cite{Bamba:2011ky}. 
The paper is organized as follows. 
In Sec.~\ref{aba:sec2}, 
we explicitly show that the future crossings of the phantom 
divide line $w_{\mathrm{DE}}=-1$ are the generic feature in the existing 
viable $f(R)$ gravity models.  
In Sec.~\ref{aba:sec3}, 
we demonstrate that the crossing of the phantom divide 
can be realized in the combined $f(T)$ theory 
having the exponential and logarithmic terms. 
In Sec.~\ref{aba:sec4}, 
we evaluate the effective EoS for the universe 
when a finite-time future singularity occurs 
in non-local gravity. The effective EoS corresponds to the ratio of 
the total pressure to the total energy density of the universe. 
We use units of $k_\mathrm{B} = c = \hbar = 1$ and denote the
gravitational constant $8 \pi G$ by
${\kappa}^2 \equiv 8\pi/{M_{\mathrm{Pl}}}^2$
with the Planck mass of $M_{\mathrm{Pl}} = G^{-1/2} = 1.2 \times 10^{19}$GeV.

\section{Future crossing of the phantom divide in $f(R)$ gravity}\label{aba:sec2} 

The action of $f(R)$ gravity 
is given by
%
$
I_{f(R)} = \int d^4 x \sqrt{-g} f(R)/\left(2\kappa^2\right)
$, 
%
where $g$ is the determinant of the metric tensor $g_{\mu\nu}$. 
Here, we use the standard metric formalism. 
It is known that viability conditions for $f(R)$ gravity are 
(a) positivity of the effective gravitational coupling, 
(b) stability of cosmological perturbations\cite{Nojiri:2003ft, 
Dolgov:2003px}, 
(c) asymptotic behavior to the standard $\Lambda$-Cold-Dark-Matter 
($\Lambda\mathrm{CDM}$) model in the large curvature regime, 
(d) stability of the late-time de Sitter point\cite{Muller:1987hp
}, 
(e) constraints from the equivalence principle, 
and 
(f) solar-system constraints\cite{Chiba:2003ir
}. 
%
We consider the following four viable models:  
(i) Hu-Sawicki\cite{Hu:2007nk}, 
$f_{\mathrm{HS}} \equiv R - 
\left[c_1 R_{\mathrm{HS}} \left(R/R_{\mathrm{HS}}\right)^p\right]/ 
\left[ c_2 \left(R/R_{\mathrm{HS}}\right)^p + 1\right]$, 
where $c_1$, $c_2$ $p(>0)$, $R_{\mathrm{HS}}(>0)$ are constant 
parameters (for an extended model, see Ref.~\refcite{Nojiri-Odintsov}). 
(ii) Starobinsky\cite{Starobinsky:2007hu}, 
$f_{\mathrm{S}} \equiv R + 
\lambda R_{\mathrm{S}} \left[
\left(1+R^2/R_{\mathrm{S}}^2 \right)^{-n}-1 
\right]$, 
where $\lambda (>0)$, $n (>0)$, $R_{\mathrm{S}}$ are constant 
parameters. 
(iii) Tsujikawa\cite{Tsujikawa:2007xu}, 
$f_{\mathrm{T}} \equiv 
R - \mu R_{\mathrm{T}} 
\tanh\left( R/R_{\mathrm{T}} \right)$, 
where $\mu (>0)$, $R_{\mathrm{T}} (>0)$ are constant 
parameters. 
(iv) the exponential gravity\cite{Exponential-Gravity
} models, 
$f_{\mathrm{E}} \equiv 
R -\beta R_{\mathrm{E}}\left[1-\exp\left(-R/R_{\mathrm{E}}\right)\right]$, 
where $\beta$ and $R_{\mathrm{E}}$ are constant parameters. 
It has been examined that 
the crossing of the phantom divide can be realized in 
the above viable $f(R)$ models on the past. 
We therefore explore the future evolution of $w_{\mathrm{DE}}$. 
We take the flat Friedmann-Lema\^{i}tre-Robertson-Walker (FLRW) metric 
$ds^2 = - dt^2 + a^2(t) \sum_{i=1,2,3}\left(dx^i\right)^2
$. 
In this background, $w_{\mathrm{DE}}$ is given by
\begin{equation}
w_{\mathrm{DE}} = \frac{
-\left(1/2\right) \left( FR - f \right) 
+\ddot{F}+2H \dot{F}
-\left(1-F\right)\left(2\dot{H}+3H^2\right)}{
\left(1/2\right) \left( FR - f \right) 
-3H \dot{F} 
+3\left(1-F\right)H^2 
}\,,
\label{eq:SEC2-add-1}
\end{equation}
where $F(R) \equiv d f(R)/dR$, and 
the dot denotes the time derivative of $\partial/\partial t$, 
and $\dot{H} \equiv \dot{a}/a$ is the Hubble parameter. 
As a result, it has explicitly been demonstrated that in the future 
the crossings of the phantom divide are the generic feature in the viable 
$f(R)$ gravity models (i)--(iv) shown above. 
We mention that $f(R)$ gravity models with realizing 
the crossings of the phantom divide have been reconstructed 
analytically\cite{Bamba:2008hq} and numerically\cite{Bamba:2009kc}. 
The new cosmological ingredient is that in the future 
the sign of $\dot{H}$ changes from 
negative to positive due to the dominance 
of dark energy over non-relativistic matter. 
This is a common physical phenomena to the existing viable $f(R)$ 
models and thus it is one of the peculiar properties of $f(R)$ gravity 
models characterizing the deviation from the $\Lambda\mathrm{CDM}$ model.

\section{Equation of state for dark energy in $f(T)$ theory}\label{aba:sec3} 

There is another procedure to study gravity beyond general relativity by 
using the Weitzenb\"{o}ck connection, which has no curvature but 
torsion, rather than the curvature defined by the Levi-Civita connection. 
This approach is referred to ``teleparallelism"\cite{Teleparallelism}. 
The teleparallel Lagrangian density described by the torsion scalar $T$ 
has been extended to a function of $T$\cite{Bengochea:2008gz, 
Linder:2010py} to account for the late-time cosmic acceleration as 
well as inflation in the early universe\cite{F-F}. 
This idea is equivalent to the concept of $f(R)$ gravity, in which 
the Ricci scalar $R$ in the Einstein-Hilbert action is promoted to 
a function of $R$. 
The modified teleparallel action for $f(T)$ theory 
is given by\cite{Linder:2010py} 
%
$
I_{f(T)}=1/\left(2\kappa^2\right) \int d^4x |e| \left( T + f(T) \right)
$, 
%
where $|e|= \det \left(e^A_\mu \right)=\sqrt{-g}$. 
In the teleparallelism, 
orthonormal tetrad components $e_A (x^{\mu})$ are used, 
where an index $A$ runs over $0, 1, 2, 3$ for the 
tangent space at each point $x^{\mu}$ of the manifold. 
In the flat FLRW background, by using the analysis method 
in Ref.~\refcite{Hu:2007nk}, 
we explicitly illustrate the cosmological evolution of $w_{\mathrm{DE}}$ 
in $f(T)$ gravity, 
expressed as~\cite{Linder:2010py} 
\begin{equation}
w_{\mathrm{DE}} = 
-1+\frac{T^\prime}{3T}\frac{f_T+2Tf_{TT}}{f/T-2f_T}=-\frac{f/T-f_T+2Tf_{TT}}{
\left(1+f_T+2Tf_{TT}\right)\left(f/T-2f_T\right)}\,, 
\label{eq:SEC3-add-1} 
\end{equation}
where a prime denotes a derivative with respect to $\ln a$, 
$f_T \equiv df(T)/dT$ and $f_{TT} \equiv d^2f(T)/dT^2$. 
Since we are interested in the late time universe, 
we consider only non-relativistic matter (cold dark matter and baryon), 
whose pressure is approximately zero. 
We have constructed an $f(T)$ theory by combining the logarithmic and 
exponential terms in order to realize the crossing of the phantom divide: 
$
f(T)= \gamma \left\{ T_0 \left(uT_0/T\right)^{-1/2} \ln 
\left(uT_0/T\right) 
-T \left[1-\exp\left(uT_0/T\right) \right]
\right\}
$
with
$
\gamma \equiv \left(1-\Omega_{\mathrm{m}}^{(0)}\right) /\left\{ 
2u^{-1/2}+\left[1-\left(1-2u\right) \exp\left(u\right) \right]\right\}
$, 
where $T_0$ 
is the current torsion 
and $u$ is a constant. 
Moreover, 
$\Omega_{\mathrm{m}}^{(0)} \equiv 
\rho_{\mathrm{m}}^{(0)}/\rho_{\mathrm{crit}}^{(0)}$, 
where 
$\rho_{\mathrm{m}}^{(0)}$
is the energy density of non-relativistic matter 
at the present time and 
$\rho_{\mathrm{crit}}^{(0)} = 3H_0^2/\kappa^2$ is the critical 
density with $H_0$ being the current Hubble parameter.  
We have shown that the crossing in the combined $f(T)$ theory 
is from $w_{\mathrm{DE}} > -1$ to $w_{\mathrm{DE}} < -1$, 
which is opposite to the typical manner in $f(R)$ gravity models.

\section{Effective equation of state in non-local gravity}\label{aba:sec4} 

Non-local gravity produced by quantum effects 
has been proposed in Ref.~\refcite{Deser:2007jk}.  
It is known that 
matter instability\cite{Dolgov:2003px} 
occurs in $f(R)$ gravity and 
the curvature inside matter sphere 
becomes very large and hence the curvature singularity 
could appear\cite{Arbuzova:2010iu, Bamba:2011sm}. 
It is important to examine whether 
there exists the curvature singularity, called 
``the finite-time future singularities'', 
in non-local gravity. 
The starting action of non-local gravity is given by
\begin{equation} 
\label{eqn:O26-12_nl1}
S=\int d^4 x \sqrt{-g}\left\{
\frac{1}{2\kappa^2}\left[ R\left(1 + f(\Box^{-1}R )\right) -2 \Lambda \right]
+ \mathcal{L}_\mathrm{matter} 
\right\}
\, .
\end{equation}
Here, $g$ is the determinant of the metric tensor $g_{\mu\nu}$, 
$f$ is some function, $\Box \equiv g^{\mu \nu} {\nabla}_{\mu} {\nabla}_{\nu}$ 
with ${\nabla}_{\mu}$ being the covariant derivative 
is the covariant d'Almbertian for a scalar field, 
$\Lambda$ is a cosmological constant, 
and 
$\mathcal{L}_\mathrm{matter}$ 
is the matter Lagrangian. 
The above action in Eq.~(\ref{eqn:O26-12_nl1}) 
can be rewritten by introducing two scalar fields $\eta$ 
and $\xi$ in the following form: 
\begin{eqnarray}
\label{eqn:O26-12_nl2}
\hspace{-6mm}
&&
S = 
\int d^4 x \sqrt{-g}\left\{
\frac{1}{2\kappa^2}\left[R\left(1 + f(\eta)\right) 
 - \partial_\mu \xi \partial^\mu \eta - \xi R - 2 \Lambda \right]
+ \mathcal{L}_\mathrm{matter} 
\right\}
\, .
\end{eqnarray}
%
We take the flat Friedmann-Lema\^{i}tre-Robertson-Walker (FLRW) metric. 
We consider the case in which 
the scalar fields $\eta$ and $\xi$ only depend on time. 
In this background, by deriving 
the gravitational field equations and the equations of motion for the scalar fields $\eta$ and $\xi$ and using these equations, 
we examine whether there exists the finite-time future singularities 
in non-local gravity. 
We analyze an asymptotic solution of 
the gravitational field equations in the limit of the time when the finite-time future singularities appear. 
We consider the case in which the Hubble parameter 
$H$ is expressed as 
%
$
H \sim h_{\mathrm{s}} \left( t_{\mathrm{s}} - t 
\right)^{-q}
$, 
%
where 
$h_{\mathrm{s}}$ is a positive constant, 
$q$ is a non-zero constant larger than $-1$ 
$(q > -1, q \neq 0)$, and 
$t_{\mathrm{s}}$ is the time when the finite-time future singularity 
appears. 
We only consider the period $0< t < t_{\mathrm{s}}$ 
because $H$ should be real number.
When $t\to t_{\mathrm{s}}$, 
for $q>1$, 
$H \sim h_{\mathrm{s}} \left( t_{\mathrm{s}} - t 
\right)^{-q}$ 
as well as 
$\dot{H} \sim q h_{\mathrm{s}} \left( t_{\mathrm{s}} - t 
\right)^{-\left(q+1\right)}$ become infinity 
and hence the scalar curvature $R$ diverges. 
For $-1 < q < 0$ and $0 < q < 1$, $H$ is finite, 
but $\dot{H}$ becomes infinity and therefore $R$ also diverges. 
{}From 
$
H \sim h_{\mathrm{s}} \left( t_{\mathrm{s}} - t 
\right)^{-q}
$, 
we obtain
%
$
a \sim a_{\mathrm{s}} \exp \left\{
\left[ h_{\mathrm{s}}/\left( q-1 \right) \right] 
\left( t_{\mathrm{s}} - t 
\right)^{-\left(q-1\right)}
\right\}
$, 
%
where $a_{\mathrm{s}}$ is a constant. 
%
$\eta$ is described as 
%
%
In the limit $t\to t_{\mathrm{s}}$, 
for $q>1$, $\dot{H} \ll H^2$ and hence $R \sim 12 H^2$, 
whereas for $-1 < q < 0$ and $0 < q < 1$, 
$\dot{H} \gg H^2$ and hence $R \sim 6\dot{H}$. 
By taking the leading term in terms of $\left( t_{\mathrm{s}} - t 
\right)$, we obtain 
%
$
\eta \sim - \left[ 4h_{\mathrm{s}}/\left( q-1 \right) \right] 
\left( t_{\mathrm{s}} - t \right)^{-\left(q-1\right)} 
+ {\eta}_{\mathrm{c}}
\,\,\, 
(q > 1) 
$, 
$
\eta \sim - \left[ 6h_{\mathrm{s}}/\left( q-1 \right) \right] 
\left( t_{\mathrm{s}} - t \right)^{-\left(q-1\right)} 
+ {\eta}_{\mathrm{c}}
\,\,\, 
(-1 < q < 0\,, \, 0 < q < 1)
$, 
%
where ${\eta}_{\mathrm{c}}$ is an integration constant. 
We take a form of $f(\eta)$ as 
%
$
f(\eta) = f_{\mathrm{s}} \eta^{\sigma}
$, 
%
where $f_{\mathrm{s}} (\neq 0)$ and $\sigma (\neq 0)$ are non-zero constants. 
{}From the expression of $a$, 
we see that 
when $t\to t_{\mathrm{s}}$, 
for $q>1$, $a \to \infty$, 
whereas 
for $-1 < q < 0$ and $0 < q < 1$, 
$a \to a_{\mathrm{s}}$. 
Moreover, 
it follows from 
$
H \sim h_{\mathrm{s}} \left( t_{\mathrm{s}} - t 
\right)^{-q}
$ 
and 
$\rho_{\mathrm{eff}} = 3 H^2/\kappa^2$ 
that 
for $q>0$, $H \to \infty$ and therefore 
$\rho_{\mathrm{eff}} = 3 H^2/\kappa^2 \to \infty$, 
whereas 
for $-1 < q < 0$, $H$ asymptotically becomes finite 
and also $\rho_{\mathrm{eff}}$ asymptotically approaches a finite 
constant value $\rho_{\mathrm{s}}$. 
On the other hand, from 
$\dot{H} \sim q h_{\mathrm{s}} \left( t_{\mathrm{s}} - t 
\right)^{-\left(q+1\right)}$ and 
$P_{\mathrm{eff}} = - \left( 2\dot H + 3H^2 \right)/\kappa^2$ 
we find that 
for $q>-1$, $\dot{H} \to \infty$ and hence 
$P_{\mathrm{eff}} = -\left(2\dot H + 3H^2\right)/\kappa^2 \to \infty$. 
Here, 
$\rho_{\mathrm{eff}}$ and $P_{\mathrm{eff}}$ are 
the effective energy density and pressure of the universe, respectively. 
It is known that 
the finite-time future singularities can be classified 
in the following manner\cite{Nojiri:2005sx}: 
%
%
Type I (``Big Rip''~\cite{Big-Rip}):\ 
In the limit $t\to t_{\mathrm{s}}$, 
$a \to \infty$,
$\rho_{\mathrm{eff}} \to \infty$ and
$| P_{\mathrm{eff}} | \to \infty$. 
The case in which 
$\rho_\mathrm{{eff}}$ and $P_{\mathrm{eff}}$ becomes finite values 
at $t = t_{\mathrm{s}}$ 
is also included. 
%
%
Type II (``sudden''~\cite{sudden}):\ 
In the limit $t\to t_{\mathrm{s}}$, 
$a \to a_{\mathrm{s}}$, 
$\rho_{\mathrm{eff}} \to \rho_{\mathrm{s}}$ and 
$| P_{\mathrm{eff}} | \to \infty$. 
%
%
Type III:\ 
In the limit $t\to t_{\mathrm{s}}$, 
$a \to a_{\mathrm{s}}$, 
$\rho_{\mathrm{eff}} \to \infty$ and
$| P_{\mathrm{eff}} | \to \infty$. 
%
%
Type IV:\ 
In the limit $t\to t_{\mathrm{s}}$, 
$a \to a_{\mathrm{s}}$, 
$\rho_{\mathrm{eff}} \to 0$, 
$| P_{\mathrm{eff}} | \to 0$, 
and higher derivatives of $H$ diverge. 
The case in which $\rho_{\mathrm{eff}}$ and/or $| P_{\mathrm{eff}} |$ 
asymptotically approach finite values is also included. 
%
%
For $q>1$, the Type I (``Big Rip'') singularity, 
for $0<q<1$, the Type III singularity, 
and for $-1<q<0$, the Type II (``sudden'') singularity. 
If ${\eta}_{\mathrm{c}} \neq 0$ and ${\xi}_{\mathrm{c}} = 1$, 
in a model with $\sigma < 0$, 
there can exist the finite-time future singularities 
with the property of 
the Type I (``Big Rip'') singularity for $q > 1$. 
If ${\eta}_{\mathrm{c}} \neq 0$, 
in a model with satisfying the condition 
$f_{\mathrm{s}} {\eta}_{\mathrm{c}}^{\sigma-1} 
\left( 6\sigma - {\eta}_{\mathrm{c}} 
\right) + {\xi}_{\mathrm{c}} -1 = 0$, 
there can exist the finite-time future singularities 
with the property of 
the Type III singularity for $0 < q < 1$ 
and that of the Type II (``sudden'') singularity for $-1 < q < 0$. 
In the special case of ${\eta}_{\mathrm{c}}= 0$, 
the finite-time future singularities 
described by 
$
H \sim h_{\mathrm{s}} \left( t_{\mathrm{s}} - t 
\right)^{-q}
$ 
cannot occur. 
The limit on a constant EoS for dark energy 
in a flat universe has been estimated as 
$w_{\mathrm{DE}} = -1.10 \pm 0.14 \, 
(68 \% \, \mathrm{CL})$ 
by combining the data of 
Seven-Year Wilkinson Microwave Anisotropy Probe (WMAP) 
Observations~\cite{Komatsu:2010fb} 
with the latest distance measurements from the BAO in the 
distribution of galaxies and the Hubble constant measurement. 
We estimate the present value of $w_{\mathrm{eff}}$. 
Here, 
we regard $w_{\mathrm{eff}}$ as being approximately equal to $w_{\mathrm{DE}}$ 
at the present time ($w_{\mathrm{eff}} \approx w_{\mathrm{DE}}$) 
because the energy density of dark energy is dominant over that of 
non-relativistic matter at the present time. 
For 
$q > 1$ with $\sigma < 0$, 
we take 
$\sigma = -1$, $q = 2$, $h_{\mathrm{s}} = 1 \, [\mathrm{GeV}]^{-1}$ 
and $t_{\mathrm{s}} = 2 t_{\mathrm{p}}$, 
where $t_{\mathrm{p}}$ is the present time. 
The current value of the Hubble parameter is given by  
$H_{\mathrm{p}} = 2.1 h \times 10^{-42} \mathrm{GeV}$\cite{Kolb-and-Turner}
with $h = 0.7$\cite{Komatsu:2010fb, Freedman:2000cf}. 
In this case, 
if $f_{\mathrm{s}} = -3.0 \times 10^{-43}$, $w_{\mathrm{eff}} = -1.10$, and 
if $f_{\mathrm{s}} = -2.1 \times 10^{-43}$, $w_{\mathrm{eff}} = -0.93$. 
For $0 < q < 1$, 
we take 
$\sigma = 1$, $q = 1/2$, $h_{\mathrm{s}} = 1 \, [\mathrm{GeV}]^{1/2}$, 
${\eta}_{\mathrm{c}} = 1$ and $t_{\mathrm{s}} = 2 t_{\mathrm{p}}$. 
In this case, 
if $f_{\mathrm{s}} = 7.9 \times 10^{-2}$, $w_{\mathrm{eff}} = -1.10$, and 
if $f_{\mathrm{s}} = 6.6 \times 10^{-2}$, $w_{\mathrm{eff}} = -0.93$. 
For $-1 < q < 0$, we have $w_{\mathrm{eff}} > 0$. 
Thus, 
the present observed value of $w_{\mathrm{DE}}$\cite{Komatsu:2010fb}
can be realized in our models. 

\section{Summary}\label{aba:sec5} 

We have investigated the EoS for dark energy 
$w_{\mathrm{DE}}$ in $f(R)$ gravity as well as $f(T)$ theory. 
We have shown that the future crossings of the phantom divide line are 
the generic feature in the existing viable $f(R)$ gravity models. 
It has also been demonstrated that the crossing of the phantom divide line can 
be realized in an $f(T)$ theory constructed by combining 
the exponential and logarithmic terms. 
Furthermore, we have explored the effective EoS for the universe 
when the finite-time future singularities occur in non-local gravity.

\section*{Acknowledgments}

The author thanks Professor Chao-Qiang Geng, 
Dr. Chung-Chi Lee, Dr. Ling-Wei Luo, 
Professor Shin'ichi Nojiri, Professor Sergei D. Odintsov 
and Professor Misao Sasaki 
for their collaborations 
in our works~\cite{Bamba:2010iy, BGLL-BGL, 
Bamba:2011ky} very much.



\begin{thebibliography}{99}


\bibitem{SN1}
%
 S.~Perlmutter {\it et al.}  [SNCP Collaboration],
 Astrophys.\ J.\  {\bf 517}, 565 (1999);\  
%
  A.~G.~Riess {\it et al.}  [Supernova Search Team Collaboration],
  Astron.\ J.\  {\bf 116}, 1009 (1998). 
%

\bibitem{WMAP}
%
D.~N.~Spergel {\it et al.}  [WMAP Collaboration],
Astrophys.\ J.\ Suppl.\  {\bf 148}, 175 (2003);\ 
%
 Astrophys.\ J.\ Suppl.\  {\bf 170}, 377 (2007);\ 
%
 E.~Komatsu {\it et al.}  [WMAP Collaboration],
 Astrophys.\ J.\ Suppl.\  {\bf 180}, 330 (2009). 
%

\bibitem{Komatsu:2010fb}
E.~Komatsu {\it et al.}  [WMAP Collaboration],
Astrophys.\ J.\ Suppl.\ {\bf 192}, 18 (2011).

\bibitem{LSS}
%
  M.~Tegmark {\it et al.}  [SDSS Collaboration],
  Phys.\ Rev.\  D {\bf 69}, 103501 (2004);\ 
%
  U.~Seljak {\it et al.}  [SDSS Collaboration],
  Phys.\ Rev.\  D {\bf 71}, 103515 (2005).
%

\bibitem{Eisenstein:2005su}
  D.~J.~Eisenstein {\it et al.}  [SDSS Collaboration],
  Astrophys.\ J.\  {\bf 633}, 560 (2005).

\bibitem{Jain:2003tba}
  B.~Jain and A.~Taylor,
  Phys.\ Rev.\ Lett.\  {\bf 91}, 141302 (2003).


\bibitem{Review-Nojiri-Odintsov}
%
  S.~Nojiri and S.~D.~Odintsov,
  Phys.\ Rept.\  {\bf 505}, 59 (2011);\ 
%
 eConf {\bf C0602061}, 06 (2006)
 [Int.\ J.\ Geom.\ Meth.\ Mod.\ Phys.\  {\bf 4}, 115 (2007)]
 [arXiv:hep-th/0601213].

\bibitem{Book-Capozziello-Faraoni}
S.~Capozziello and V.~Faraoni,
\textit{Beyond Einstein Gravity}
(Springer, 2010). 


\bibitem{observational-status}
%
  U.~Alam, V.~Sahni and A.~A.~Starobinsky,
  JCAP {\bf 0406}, 008 (2004);\ 
%
  JCAP {\bf 0702}, 011 (2007);\ 
%
  S.~Nesseris and L.~Perivolaropoulos, 
  JCAP {\bf 0701}, 018 (2007).
%
 
\bibitem{Bamba:2010iy}
  K.~Bamba, C.~Q.~Geng and C.~C.~Lee,
  JCAP {\bf 1011}, 001 (2010).

\bibitem{BGLL-BGL}
%
  K.~Bamba, C.~Q.~Geng, C.~C.~Lee and L.~W.~Luo,
  JCAP {\bf 1101}, 021 (2011);\ 
%
  K.~Bamba, C.~Q.~Geng and C.~C.~Lee,
  arXiv:1008.4036 [astro-ph.CO].

\bibitem{Bamba:2011ky}
  K.~Bamba, S.~Nojiri, S.~D.~Odintsov and M.~Sasaki,
  arXiv:1104.2692 [hep-th], 
to be published in General Relativity and Gravitation. 


%
\bibitem{Nojiri:2003ft}
  S.~Nojiri and S.~D.~Odintsov,
  Phys.\ Rev.\  D {\bf 68}, 123512 (2003). 

\bibitem{Dolgov:2003px}
  A.~D.~Dolgov and M.~Kawasaki,
  Phys.\ Lett.\  B {\bf 573}, 1 (2003). 

%
\bibitem{Muller:1987hp}
  V.~Muller, H.~J.~Schmidt and A.~A.~Starobinsky,
  Phys.\ Lett.\  B {\bf 202}, 198 (1988).

%
\bibitem{Chiba:2003ir}
  T.~Chiba,
  Phys.\ Lett.\  B {\bf 575}, 1 (2003).


\bibitem{Hu:2007nk}
  W.~Hu and I.~Sawicki,
  Phys.\ Rev.\  D {\bf 76}, 064004 (2007). 

\bibitem{Nojiri-Odintsov}
%
  S.~Nojiri and S.~D.~Odintsov,
  Phys.\ Lett.\  B {\bf 657}, 238 (2007);\ 
%
  Phys.\ Rev.\  D {\bf 77}, 026007 (2008).
%

\bibitem{Starobinsky:2007hu}
  A.~A.~Starobinsky,
  JETP Lett.\  {\bf 86}, 157 (2007).

\bibitem{Tsujikawa:2007xu}
  S.~Tsujikawa,
  Phys.\ Rev.\  D {\bf 77}, 023507 (2008).

\bibitem{Exponential-Gravity}
%
  G.~Cognola, E.~Elizalde, S.~Nojiri, S.~D.~Odintsov, L.~Sebastiani and 
S.~Zerbini,
  Phys.\ Rev.\  D {\bf 77}, 046009 (2008);\ 
%
  E.~V.~Linder,
  Phys.\ Rev.\  D {\bf 80}, 123528 (2009);\ 
%
  K.~Bamba, C.~Q.~Geng and C.~C.~Lee,
  JCAP {\bf 1008}, 021 (2010).
%

\bibitem{Bamba:2008hq}
  K.~Bamba, C.~Q.~Geng, S.~Nojiri and S.~D.~Odintsov,
  Phys.\ Rev.\  D {\bf 79}, 083014 (2009). 

\bibitem{Bamba:2009kc}
  K.~Bamba and C.~Q.~Geng,
  Prog.\ Theor.\ Phys.\  {\bf 122}, 1267 (2009). 


\bibitem{Teleparallelism}
%
  F.~W.~Hehl, P.~Von Der Heyde, G.~D.~Kerlick and J.~M.~Nester,
  Rev.\ Mod.\ Phys.\  {\bf 48}, 393 (1976);\ 
%
  K.~Hayashi and T.~Shirafuji,
  Phys.\ Rev.\  D {\bf 19}, 3524 (1979)
  [Addendum-ibid.\  D {\bf 24}, 3312 (1982)].
%

\bibitem{Bengochea:2008gz}
  G.~R.~Bengochea and R.~Ferraro,
  Phys.\ Rev.\  D {\bf 79}, 124019 (2009).

\bibitem{Linder:2010py}
  E.~V.~Linder,
  Phys.\ Rev.\  D {\bf 81}, 127301 (2010)
  [Erratum-ibid.\  D {\bf 82}, 109902 (2010)].

\bibitem{F-F}
%
  R.~Ferraro and F.~Fiorini,
  Phys.\ Rev.\  D {\bf 75}, 084031 (2007);\ 
%
  Phys.\ Rev.\  D {\bf 78}, 124019 (2008). 
%

\bibitem{Deser:2007jk}
  S.~Deser and R.~P.~Woodard,
  Phys.\ Rev.\ Lett.\  {\bf 99}, 111301 (2007). 


\bibitem{Arbuzova:2010iu}
  E.~V.~Arbuzova and A.~D.~Dolgov,
  Phys.\ Lett.\  B {\bf 700}, 289 (2011). 

\bibitem{Bamba:2011sm}
  K.~Bamba, S.~Nojiri and S.~D.~Odintsov,
  Phys.\ Lett.\  B {\bf 698}, 451 (2011). 

\bibitem{Nojiri:2005sx}
  S.~Nojiri, S.~D.~Odintsov and S.~Tsujikawa,
  Phys.\ Rev.\  D {\bf 71}, 063004 (2005). 

\bibitem{Big-Rip}
%
  R.~R.~Caldwell, M.~Kamionkowski and N.~N.~Weinberg,
  Phys.\ Rev.\ Lett.\  {\bf 91}, 071301 (2003);\ 
%
B.~McInnes,
JHEP {\bf 0208} (2002) 029. 
%

\bibitem{sudden}
%
 J.~D.~Barrow,
 Class.\ Quant.\ Grav.\  {\bf 21}, L79 (2004);\ 
%
 S.~Nojiri and S.~D.~Odintsov,
 Phys.\ Lett.\  B {\bf 595}, 1 (2004). 
%

\bibitem{Kolb-and-Turner}
E.~W.~Kolb and M.~S.~Turner,
\textit{The Early Universe}
(Addison-Wesley, Redwood City, California, 1990).

\bibitem{Freedman:2000cf}
 W.~L.~Freedman {\it et al.}  [HST Collaboration],
 Astrophys.\ J.\  {\bf 553}, 47 (2001). 

\end{thebibliography}
\end{document}